\documentclass[traditabstract]{aa}

\usepackage{txfonts}
\usepackage{natbib,graphicx,amssymb}
\usepackage[english]{babel}
\def\revised{}

\begin{document}

\title{The effects of star formation on the low-metallicity ISM:\\
  NGC4214 mapped with \textit{Herschel}\thanks{Herschel is an ESA
    space observatory with science instruments provided by
    European-led Principal Investigator consortia and with important
    participation from NASA.}/PACS spectroscopy}

\author{
  D.~Cormier\inst{1}
  \and S.~C.~Madden\inst{1}
  \and S.~Hony\inst{1}
  \and A.~Contursi\inst{2}
  \and A.~Poglitsch\inst{2}
  \and F.~Galliano\inst{1}
  \and E.~Sturm\inst{2}
  \and V.~Doublier\inst{2}
  \and H.~Feuchtgruber\inst{2}
  \and M.~Galametz\inst{1}
  \and N.~Geis\inst{2}
  \and J.~de Jong\inst{2}
  \and K.~Okumura\inst{1}
  \and P.~Panuzzo\inst{1}
  \and M.~Sauvage\inst{1}
}

\institute{ Laboratoire AIM, CEA/DSM - CNRS - Universit\'e Paris
  Diderot, Irfu/Service d'Astrophysique, CEA Saclay, 91191
  Gif-sur-Yvette, France \email{diane.cormier@cea.fr} \and
  Max-Planck-Institut fuer extraterrestrische Physik, Postfach 1312,
  85741 Garching, Germany}


\abstract{ We present {\it Herschel}/PACS spectroscopic maps of the
  dwarf galaxy NC4214 observed in 6 far infrared fine-structure lines:
  [C~{\sc ii}] 158$\mu$m, [O~{\sc iii}] 88$\mu$m, [O~{\sc i}]
  63$\mu$m, [O~{\sc i}] 146$\mu$m, [N~{\sc ii}] 122$\mu$m, and [N~{\sc
    ii}] 205$\mu$m. The maps are sampled to the full telescope spatial
  resolution and reveal unprecedented detail on $\sim$ {{\revised 150}} pc size
  scales. We detect [C~{\sc ii}] emission over the whole mapped area,
  [O~{\sc iii}] being the most luminous FIR line. The ratio of [O~{\sc
    iii}]/[C~{\sc ii}] {{\revised peaks at about 2 toward the sites of massive
    star formation}}, higher than ratios seen in dusty starburst
  galaxies. The [C~{\sc ii}]/CO ratios are 20 000 to 70 000 toward the
  2 massive clusters, which are at least an order of magnitude larger than
  spiral or dusty starbursts, {{\revised and cannot be reconciled with 
  single-slab PDR models}}. Toward the 2 massive star-forming
  regions, we find that L$_{\rm [CII]}$ is 0.5 to 0.8\% of the L$_{\rm TIR}$.
  All of the lines together contribute up to 2\% of L$_{\rm TIR}$. 
  These extreme findings are a consequence of the
  lower metallicity and young, massive-star formation commonly
  found in dwarf galaxies. {{\revised These conditions promote}} 
  large-scale photodissociation into the molecular reservoir, 
  {{\revised which is evident in the FIR line ratios}}. 
  This illustrates the necessity  
  {{\revised to move to multiphase models}} 
  applicable to star-forming clusters or galaxies as a whole. }

\titlerunning{The effects of star formation on the low-metallicity ISM}
\authorrunning{Cormier et al.}
\maketitle

\section{Introduction}
As building blocks of larger galaxies, dwarf galaxies in the early
universe presumably played an important role in the evolution of
galaxies we see today. Understanding how stars form and evolve under
low metallicity conditions and the process of subsequent enrichment of
the interstellar medium (ISM) will help us constrain possible scenarios
of galaxy formation and evolution. The Local Group dwarf galaxies
provide laboratories to study in detail the effects of the lower
metal abundance on the process of star formation and the feedback on
the ISM.

{\revised To examine the interplay between massive star formation and 
low metallicity ISM}, we have mapped the far infrared (FIR) fine
structure lines in the Local Group Magellanic-type irregular galaxy,
\object{NGC4214}, as part of the {\it Herschel} key proposal, SHINING
(P.I. E. Sturm). NGC4214 harbors 2 main star-forming complexes
containing hundreds of O stars as well as a super star cluster (SSC).
Our aim here is to characterise the physical conditions and the
structure of the different phases of the ISM in NGC4214. 
The proximity of NGC4214 \citep[2.9Mpc,][]{maiz-apellaniz-2002} and
its low-metallicity
\citep[$\rm{log(O/H)+12=8.2}$,][]{kobulnicky-skillman-1996} make it a
choice target. The {PACS} spectrometer onboard the {\it Herschel Space
Observatory} allows us to
zoom into the low-metallicity photodissociation regions 
(PDRs) and their surroundings, tracing
physical properties from {\revised 125 to 165} pc size scales.

The tools for this study are the FIR fine-structure lines arising from
some of the most abundant species C, O, and N, which provide the most
important cooling channels in the neutral and diffuse ionized
ISM of galaxies. These lines trace H~{\sc ii}
~regions and PDRs, which are neutral atomic
and molecular regions where penetrating FUV photons (6 eV $< h\nu <$
13.6 eV) dominate the energy balance and the chemical composition of
the gas. Most of the ISM in galaxies is contained in PDRs. The FIR
lines in combination with PDR models
\citep{tielens-hollenbach-1985,wolfire-1990,ferland-1998,kaufman-1999,abel-2005,roellig-2006}
can unveil the structure of these regions, which depends on both the
strength of the FUV radiation field density ($\rm{G_{0}}$)
illuminating the cloud and the hydrogen density ($\rm{n_{H}}$).

\section{Observations and data reduction}
We mapped the 6 FIR fine-structure lines, [C~{\sc ii}] 158, [O~{\sc
  iii}] 88, [O~{\sc i}] 63, [O~{\sc i}] 146, [N~{\sc ii}] 122, and
[N~{\sc ii}] 205$\mu$m, using the {PACS} spectrometer
\citep{poglitsch_special_issue} on {\it Herschel} 
\citep{pilbratt_special_issue}. The {PACS} array consists
of 5 x 5 spatial pixels of 9.4$^{\prime\prime}$ size offering a total
field-of-view of 47$^{\prime\prime}$ on the sky. Each spatial pixel
covers 16 spectral elements with a spectral resolution ranging from
100 to 300\,km/s. The lines are not spectrally resolved for
NGC4214. The 6 lines were observed in raster mapping mode over a total
of 15h during the Science Demonstration Phase. The [O~{\sc iii}]
88$\mu$m and [O~{\sc i}] 63$\mu$m (``blue'') were mapped with  {\revised 5 x 5}
rasters and the other lines (``red'') with  {\revised 3 x 3} rasters, all centered on
12h15m39.08s,+36d19m35.9s (J2000). Each raster {\revised pointing} is separated by
approximately half of the array (22$^{\prime\prime}$,
24$^{\prime\prime}$) in the ``red'' range and approximately one third of 
the array (14.5$^{\prime\prime}$, 16$^{\prime\prime}$) in the ``blue''
range, giving homogeneous coverage of the galaxy and a total field of
view of 1.6$^{\prime}$x1.6$^{\prime}$. We selected the chop-nod mode
with a chop throw of 6$^{\prime}$ off the source.

\begin{figure}
\centering
\includegraphics[trim=10mm 0 0 0,width=8cm,height=10cm]{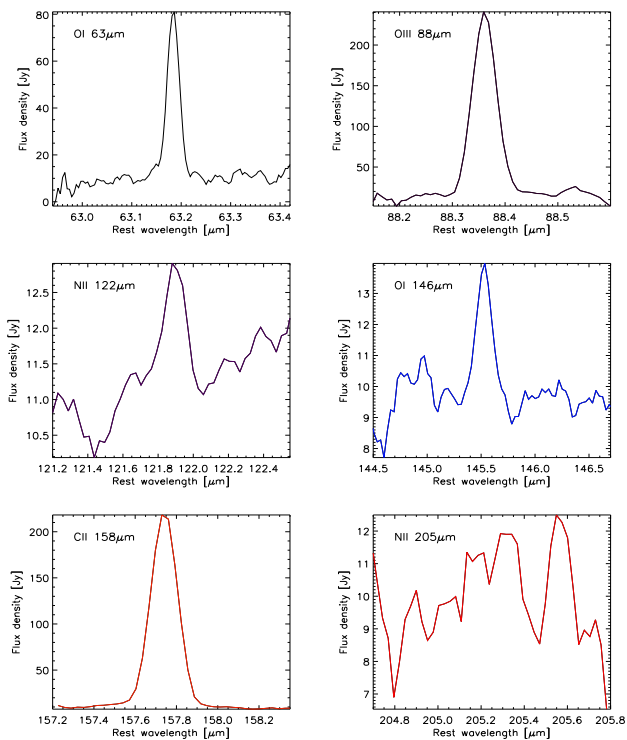}
\caption{Individual PACS spectral lines added up over the central raster.
The lines are not spectrally resolved.}
\label{fig:fig1}
\end{figure}

The data were reduced with the {PACS} spectrometer pipeline of the
{\it Herschel} Interactive Processing Environment (HIPE) v3.0.455. 
We show the 6 individual lines in Fig.~\ref{fig:fig1}. We
fitted a spline plus Gaussian to the line and baseline and measured
the $1\sigma$ noise of the residual. In this way, we estimate the
overall noise in the maps to be 0.17, 0.42, 0.27, 0.07, 0.06, and 0.30
Jy/pix for the [C~{\sc ii}] 158$\mu$m, [O~{\sc iii}] 88$\mu$m, [O~{\sc
  i}] 63$\mu$m, [O~{\sc i}] 146$\mu$m, [N~{\sc ii}] 122$\mu$m, and
[N~{\sc ii}] 205$\mu$m lines, respectively. We applied a flat field
correction and divided by calibration correction factors of 1.1
(``red'') and 1.3 (``blue'') to obtain the final reduced maps.

The rasters are combined by a drizzle scheme to produce final maps with
4$^{\prime\prime}$ {\revised grid pixel size} for the [O~{\sc iii}] and
[O~{\sc i}] 63$\mu$m lines, and 6$^{\prime\prime}$ for the other lines. To
compare the lines and create ratio maps, we scaled all maps down to
the coarser resolution of 6$^{\prime\prime}$. We analyze separately
the whole map as well as three specific regions corresponding to
NGC4214-I (center), NGC4214-II (southeast), and NGC4214-III
(northwest). {\revised These are represented by the red circles
on Fig.~\ref{fig:fig2}a. Regions I and II are centered on the [C~{\sc ii}] peaks, 
with radii encompassing most of the emisson (at the level of $\sim$50\% of the peak), 
while region III was chosen by taking into account the peak of the CO emission 
as well as the brightest counterpart of the [C~{\sc ii}] emission
($\rm{S/N} \gtrsim 15$), which is truncated by the map edge.}

\section{Results}
\begin{figure*}
\centering
\includegraphics[width=8cm,height=6cm,clip,trim=0 0 6mm 8mm]{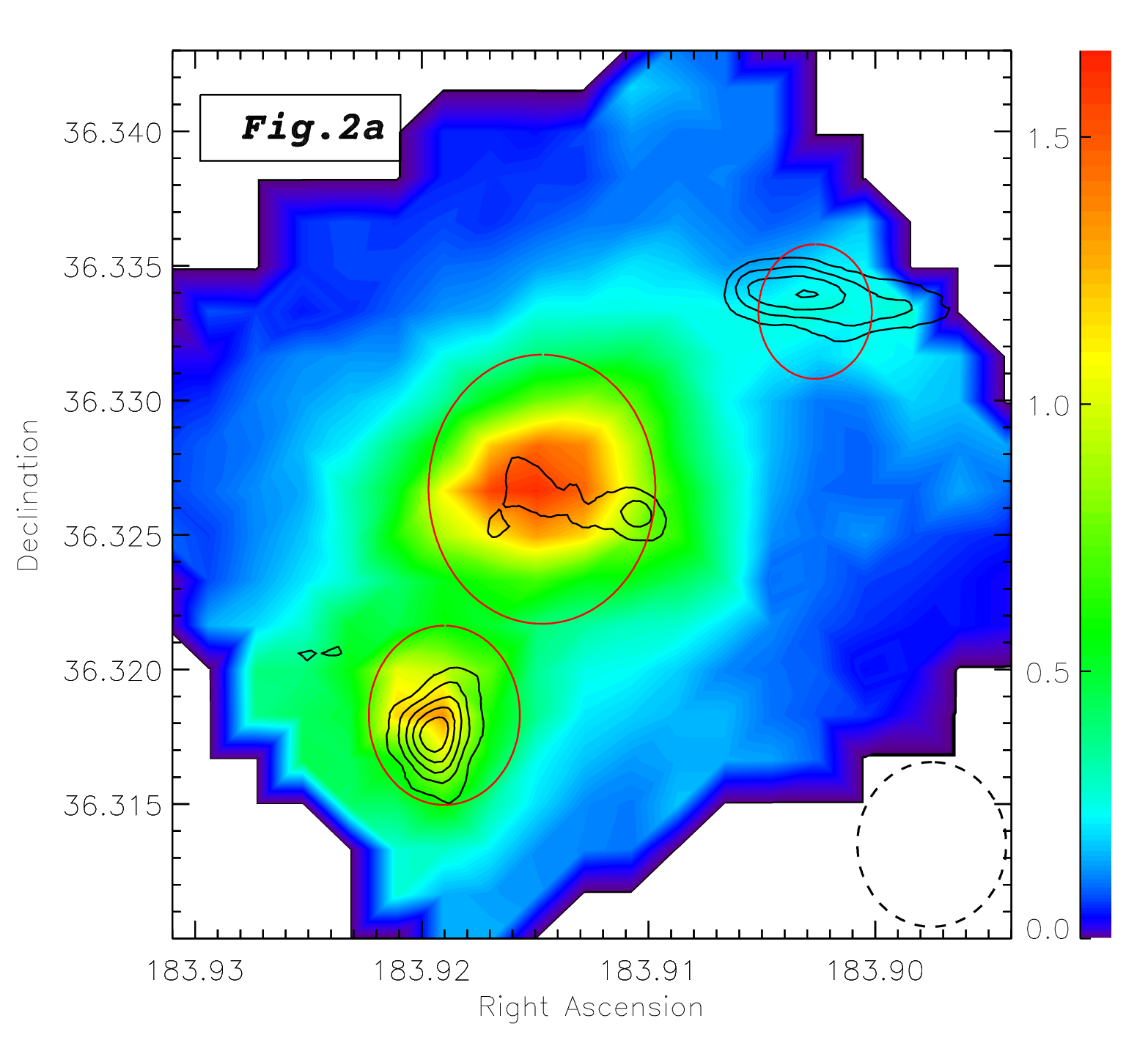}
\includegraphics[width=8cm,height=6cm,clip,trim=0 0 6mm 8mm]{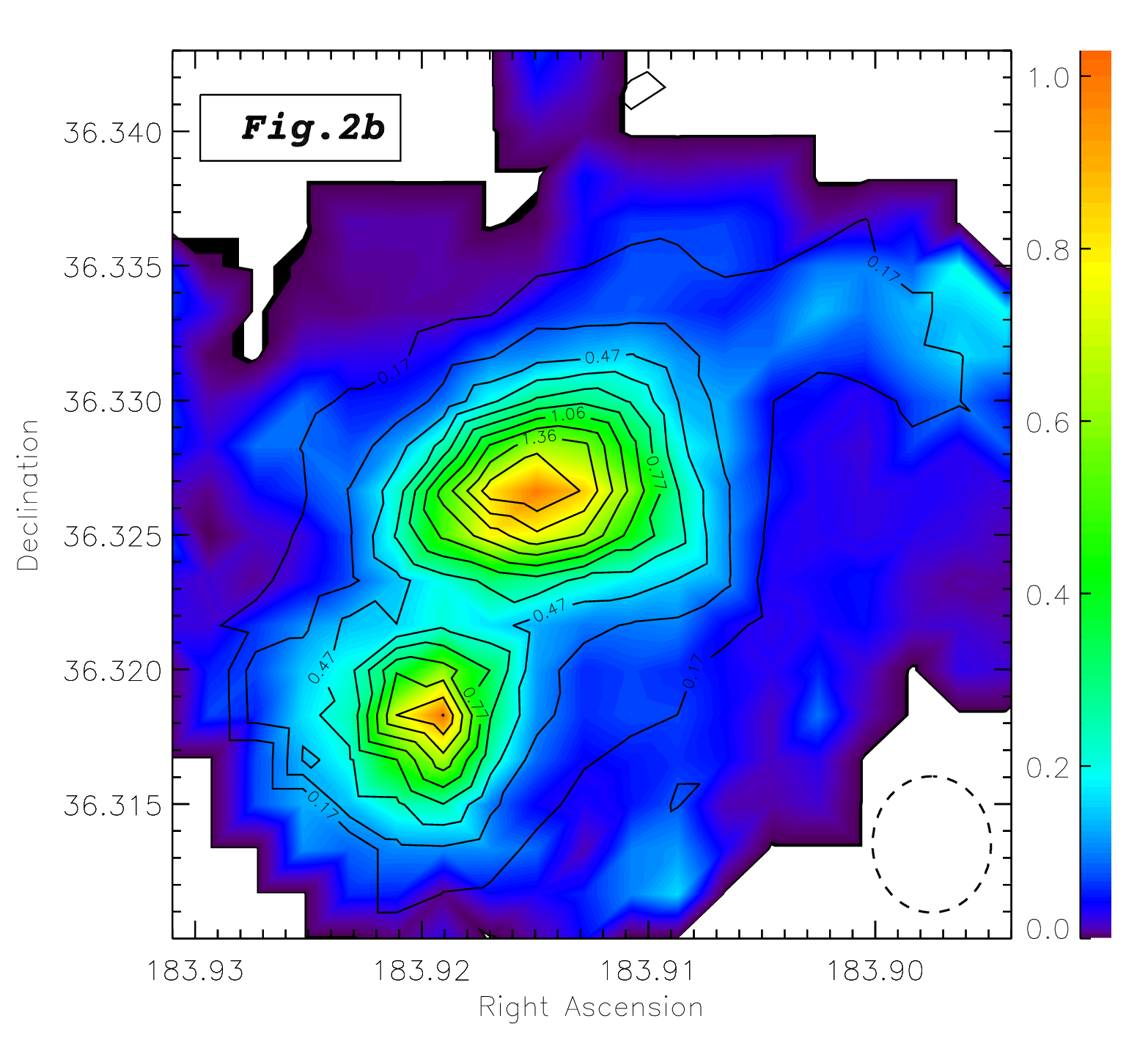}
\includegraphics[width=8cm,height=6cm,clip,trim=0 0 6mm 8mm]{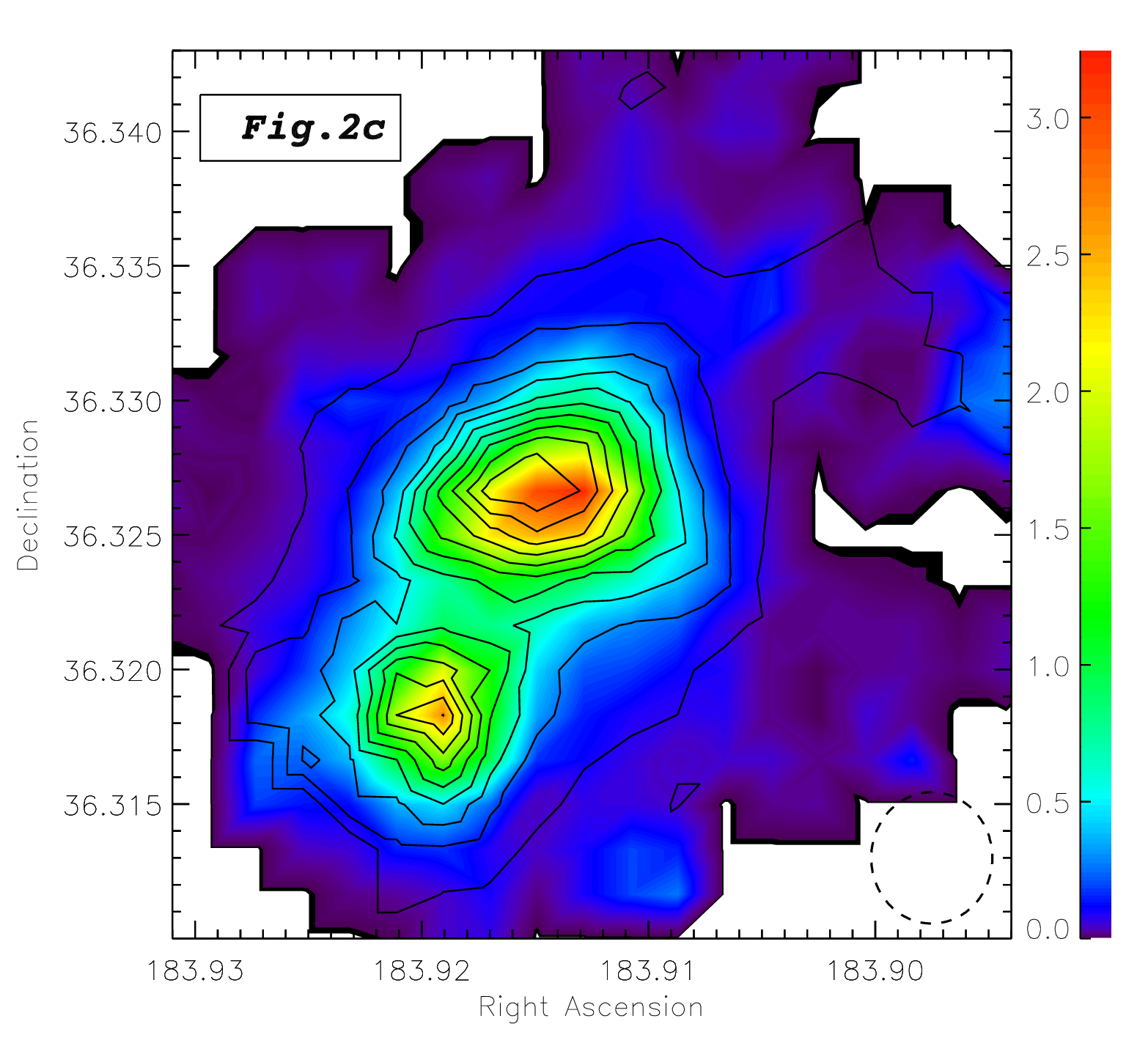}
\includegraphics[width=8cm,height=6cm,clip,trim=0 0 6mm 8mm]{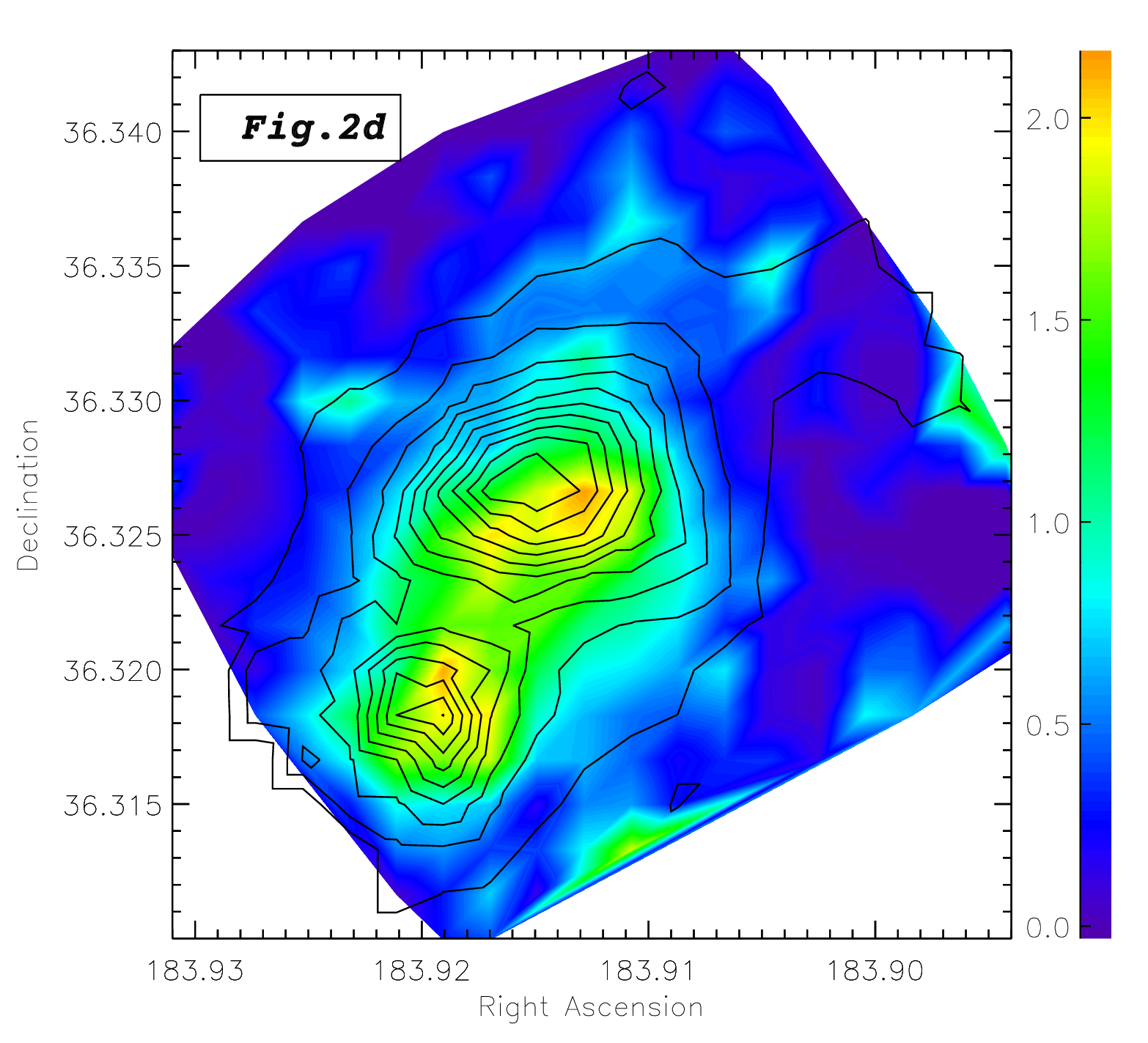}

\caption{ {\revised 
  PACS spectral maps of NGC4214.
\textit{Fig 2a:} }
  {PACS} map of the [C~{\sc
    ii}] 158$\mu$m line with CO(1-0) contours \citep{walter-2001}. The
  total field of view is 1.6$^{\prime}$x1.6$^{\prime}$ with a grid
  resolution of 6$^{\prime\prime}$. The red circles show the positions
  and sizes of the 3 regions we study in more detail: NGC4214-I
  (center - 36$^{\prime\prime}$ diameter), NGC4214-II
  (southeast-24$^{\prime\prime}$), and NGC4214-III
  (northwest-18$^{\prime\prime}$). 
  The dotted circle on the lower right corner represents the 
  {\revised PACS beam size of 9.4$^{\prime\prime}$x 9.1$^{\prime\prime}$ at 62$\mu$m, 
  9.6$^{\prime\prime}$x 8.8$^{\prime\prime}$ at 90$\mu$m, and 
  11.8$^{\prime\prime}$x 11.0$^{\prime\prime}$ at 154$\mu$m}. 
  \textit{Fig 2b:} {PACS} map of the [O~{\sc
    i}] 63$\mu$m line with [C~{\sc ii}] 158$\mu$m contours.
  \textit{Fig 2c:} {PACS} map of the [O~{\sc
    iii}] 88$\mu$m line with [C~{\sc ii}] 158$\mu$m contours. 
  \textit{Fig 2d:} Ratio map
  of [O~{\sc iii}]~88$\mu$m/[C~{\sc ii}] with [C~{\sc ii}] contours.
  {\revised Units for the color bars are $10^{-16}$ W m$^{-2}$ grid pix$^{-1}$.
  The [C~{\sc ii}]  contours begin at $0.17\times10^{-16}$ W m$^{-2}$ (above $15\sigma$) 
  and are linearly spaced up to the peak of  $1.66\times10^{-16}$ W m$^{-2}$. }}

\label{fig:fig2}
\end{figure*}

\begin{table}
  \caption{FIR line ratios}
  \begin{tabular}{@{\, }l@{\,   }c@{\,     }c@{\,     }c@{\,     }c@{\, }} 
    \hline\hline
    NGC4214 & total &  I &  II &III \\
    \hline
     RA [J2000] & 12\,15\,39.1 & 12\,15\,39.1 & 12\,15\,40.5 & 12\,15\,36.7 \\
     Dec [J2000] & $+$36\,19\,35.9 & $+$36\,19\,35.9 & $+$36\,19\,05.9 & $+$36\,20\,01.0 \\
     Aperture ($^{\prime\prime}$) & - & 18 & 12 & 9 \\
    \hline
   \multicolumn{1}{l}{Quantity} & 
    \multicolumn{3}{c}{Line ratios}\\     
    \hline
    $f_{\rm [OIII]} / f_{\rm [CII]}$  & 0.996 & 1.40 & 1.71 & 0.264 \\
    & (0.009) & (0.007) & (0.014) & (0.053) \\
    $f_{\rm{[OI]}\lambda146}/f_{\rm{[OI]}\lambda63}$& 0.058 & 0.060 & 0.072 & - \\
    & (0.015) & (0.011) & (0.023) & (0.095) \\
    $f_{{\rm [OI]} \lambda 63}/f_{\rm [CII]}$ & 0.445 & 0.474 & 0.560 & 0.438 \\
    & (0.007) & (0.004) & (0.008) & (0.035) \\
    $f_{{\rm [OI]} \lambda 146}/f_{\rm [CII]}$ & 0.026 & 0.028 & 0.040 & - \\
    & (0.007) & (0.005) & (0.013) & (0.042) \\
    $L_{\rm TIR}$ [$10^{7}~L_{\sun}$] & 37.7 & 8.60 & 4.52 & 0.522 \\
    $10^{-3} (L_{{\rm [OI]} \lambda63}$ & 2.77 & 8.11 & 6.50& 11.1 \\
    	+  $L_{\rm [CII]}$) / $L_{\rm TIR}$ \\
    $10^{-3} L_{\rm [CII]}$/$L_{\rm TIR}$ & 5.30 & 7.93 & 5.71 & 7.70 \\
    $10^{3} L_{\rm [CII]}$/$L_{\rm CO}$   & 33.8 & 75.0 & 22.9 & 4.73 \\
    \hline
  \end{tabular}
  \label{table:table1}
\end{table}

The {\it Herschel}/{PACS} maps clearly resolve the two star-forming 
regions NGC4214-I and NGC4214-II, where most of the emission
originates. The [N~{\sc ii}] 122$\mu$m emission is rather faint and
the [N~{\sc ii}] 205$\mu$m is not detected. The most prominent lines
are the [C~{\sc ii}] 158$\mu$m, [O~{\sc iii}] 88$\mu$m, [O~{\sc i}]
63$\mu$m, and, to a lesser extent, {\revised the 
[O~{\sc i}] 146$\mu$m and [N~{\sc ii}] 122$\mu$m lines}. The emission
from the [C~{\sc ii}] line is the most extended of all of the lines
(Fig.~\ref{fig:fig2}a) and covers the entire map. The total
[C~{\sc ii}] luminosity in the {PACS} map is $2.0\times10^{6}
L_{\sun}$ with about 50\% of the emission originating in the two
bright star-forming sites and 50\% in the surrounding
medium. The [C~{\sc ii}] 158$\mu$m {\revised line is known to be} one of the most
important tracers of physical conditions in PDRs because it is so
luminous and has a low critical density for collisional excitation.

Comparing the [C~{\sc ii}] and [O~{\sc i}] 63$\mu$m line emission, we
find that {\revised the latter} is a factor of two weaker integrated over the
mapped region. These two lines show a very similar distribution 
(Fig.~\ref{fig:fig2}b), both
peaking at the 2 star-forming sites and spanning a small range of
ratios. These ratios vary by less than a factor of two across the entire map. 
The [C~{\sc ii}] line, in principle, may originate in the PDR and the diffuse
media alike. The observed resemblance between the [O~{\sc i}] 63$\mu$m and
[C~{\sc ii}] maps {\revised (Fig.~\ref{fig:fig2}b) suggests} that the [C~{\sc ii}] is dominated
by the PDR component (see also Sect.~\ref{sect:origin_cii}).

The [O~{\sc i}] 146$\mu$m line is fainter than the [O~{\sc i}] 63$\mu$m line,
and is barely detected in the outer regions of the map. Toward the 2
peaks, and in total, the ratios of the two [O~{\sc i}] lines are about
0.06 (Table~\ref{table:table1}), {\revised and do not vary much 
across the map. These ratios are insensitive to density since 
their critical densities are not very different.
For moderate densities (n $< 10^5 cm^{-3}$), this ratio is 
more indicative of the gas temperature.}

The [O~{\sc iii}] 88$\mu$m line is the brightest of all 
lines observed in NGC4214 (Fig.~\ref{fig:fig2}c). 
A comparison of the [O~{\sc iii}] line map with the $HST$ optical images
of NGC4214 \citep{ubeda-2007I}, illustrates that it coincides with the star
formation sites and exhibits a peak toward the position of the SSC in
NGC4214-I, thus tracing the ionization source. 
High ratios of [O~{\sc iii}] to [C~{\sc ii}] observed here 
(Table~\ref{table:table1}) are not commonly seen in dusty
starbursts or spiral galaxies \citep{negishi-2001,malhotra-2001}, an
example being M82 with a ratio of $\sim$ 0.7. However, high ratios
have been observed in the $ISO$ observations of dwarf galaxies and may
be explained by hot stars surrounded by optically thick PDRs
\citep{hunter-2001}. The [O~{\sc iii}] / [C~{\sc ii}]
ratio map of NGC4214 (Fig.~\ref{fig:fig2}d) exhibits peaks in intensity toward
the 2 star-forming sites but drops off rapidly away from the 
peaks. We note that the peaks of the [O~{\sc iii}] / [C~{\sc
  ii}] ratio map appear to be offset from those of the [C~{\sc ii}] by
$\sim$ 6$^{\prime\prime}$, which is larger than the expected
relative pointing uncertainty (on the order of several arcsecs).
The shift in the [O~{\sc iii}] / [C~{\sc ii}] peak, corresponding to a
shift of $\sim$ 80 pc, if real, implies that the ratio peak coincides with the SSC peak,
the probable source of the ionization. Since O$^{++}$ has an ionization 
potential of 35.1 eV, [O~{\sc
  iii}] traces the highly ionized medium around early O
stars and is thus more confined than the [C~{\sc ii}] emission.
This is consistent with the AKARI findings that the [O~{\sc
  iii}] 88$\mu$m line traces excitation sources even more accurately
than the radio continuum \citep{okada-2009}.

We show the CO contours from \citet{walter-2001} in
Fig.~\ref{fig:fig2}a. There is a striking distinction between the three
different regions {\revised in terms of [C~{\sc ii}]/CO (Table~\ref{table:table1})}. 
NGC4214-III probably hosts little star formation,
which explains its paucity in excited FIR line emission compared to
its molecular reservoir. However, even the two {\revised main} regions of active star
formation exhibit very different [C~{\sc ii}]/CO ratios. One
explanation of these high [C~{\sc ii}]/CO ratios is the presence of small
clumps of CO embedded in large PDR {\revised envelopes. 
Applying this interpretation to these}
observations would imply that the star formation in the
central region of NGC4214-I has (already) affected (photodissociated)
most of the surrounding molecular reservoir, while many of the
molecular clouds in the south eastern region are relatively unscathed.
It is interesting to remark that NGC4214-II has a younger stellar
population than NGC4214-I \citep{ubeda-2007I}.

Using {\it Spitzer} data, we estimate the total infrared luminosity
(L$_{\rm TIR}$) over the area mapped with {PACS}, following
\citet{dale-2002} (see Table~\ref{table:table1}). The [C~{\sc ii}] line alone accounts for
$\sim$~0.5\% of the L$_{\rm TIR}${\revised \footnote{To compare ratios using L$_{\rm FIR}$ 
in the literature, for dwarf galaxies \citep{hunter-2001}, L$_{\rm TIR}$ is approximately 
a factor of 2 greater than L$_{\rm FIR}$, as defined by IRAS 60 and 100$\mu$m
fluxes \citep{helou-1988}.}}, 
which is at the high end of the range of normal
galaxies and starburst galaxies \citep{stacey-1991,negishi-2001} and
more typical of the elevated ratios seen in dwarf galaxies
\citep{madden-2000,hunter-2001}. 
{\revised Toward the two bright star-forming 
regions, L$_{\rm [C II]}$/L$_{\rm TIR}$ is
similar to \object{30 Dor} in the \object{LMC}
\citep{poglitsch-1995,rubin-2009}. Likewise, L$_{\rm [O III]
}$/$L_{\rm TIR}$ either toward these 2 prominent star-forming regions or over
the entire galaxy is an order of magnitude higher than what
\citet{negishi-2001} find for normal and starburst galaxies. The FIR
lines of the 2 star-forming regions contribute all together up 
to about 2\% of their TIR flux.}

\subsection{Origin of the [C~{\sc ii}] line}
\label{sect:origin_cii}
Before we can consider PDR model solutions to the observations, we
consider the origin of the C$^+$ line, which is most often used as a
diagnostic of PDR conditions because it is a primary coolant in PDRs.
However, since the ionization potential of carbon (11.3 eV) is less
than that of hydrogen, C$^+$ can {\revised exist} in both the ionized and
neutral gas. In galactic nuclei, most of the [C~{\sc ii}] emission
arises from the PDRs \citep{stacey-1991,negishi-2001}. 
With an ionization potential of 14.53 eV, the N$^+$ emission originates only
from the H~{\sc ii} regions. Therefore we use the [N~{\sc ii}] line to
correct for a possible contribution of the ionized gas to the C$^{+}$ line.
 
The [N~{\sc ii}] 205$\mu$m line was not confidently detected and could
not be analyzed but the [N~{\sc ii}] 122$\mu$m line is detected in
the two star-forming regions. We see evidence of [N~{\sc ii}]
122$\mu$m emission over the central 3x3 pixels in each region and
average them to obtain a higher signal-to-noise ratio and measure the intensity
of the line in these two regions. The observed ratios of the [N~{\sc
  ii}] 122$\mu$m/[C~{\sc ii}] 158$\mu$m are 0.04 and 0.03 in NGC4214-I
and NGC4214-II, respectively. Following \citet{malhotra-2001}, we
estimate a maximum contribution of the ionized medium to the [C~{\sc
  ii}] emission to be a factor of 11 times the flux of the [N~{\sc
  ii}] 122$\mu$m. Therefore, at least 60\% of the [C~{\sc ii}]
emission originates in PDRs. To bound the physical conditions in
NGC4214, we model the PDR (see Sect.~\ref{sect:pdr}) assuming that all [C~{\sc ii}] 
emission is from PDRs and that only 60\% originates in PDRs.

\subsection{Results from PDR models}
\label{sect:pdr}
As a first diagnostic of PDR properties, we use the model of
\citet{kaufman-1999} by means of the web interface of the PDR
Toolbox\footnote{http://dustem.astro.umd.edu/pdrt/index.html}. This
model, {\revised which assumes solar metallicity}, 
calculates the intensity of the main PDR lines as a function of
the physical conditions, which are the incident radiation field
($G_{0}$) and the density ($n_{\rm H}$). 
We can compare our observed line intensities to the model
predictions to narrow down the range of the
physical conditions.

We use [O~{\sc i}]~63$\mu$m/[C~{\sc ii}], [O~{\sc
  i}]~146$\mu$m/[C~{\sc ii}], [O~{\sc i}]~146$\mu$m/[O~{\sc
  i}]~63$\mu$m, and the ([C~{\sc ii}]+[O~{\sc i}]~63$\mu$m)/I$_{\rm
  TIR}$ and [C~{\sc ii}]/CO (Table~\ref{table:table1}). The ratio ([C~{\sc
  ii}]+[O~{\sc i}] 63$\mu$m)/I$_{TIR}$ is interesting because
it relates the ``total'' cooling of the gas to the total cooling by
the dust, so measures the efficiency of the dust to heat the gas
(photoelectric efficiency).
The observed [C~{\sc ii}]/CO ratio disagrees significantly 
with those predicted by the best-fit solution
for all the other line ratios. This has been noted before for
other low metallicity PDRs (see Sect.~\ref{sec:c-sc-ii}) and {\revised may}
be a geometry or low metallicity effect not
taken into account in the model. 
{\revised Using all the above line ratios except [C~{\sc ii}]/CO, 
the maximum values found 
for NGC4214-I are $G_{0}\footnote {in units of
  Habing field: $1.6\times10^{-3}$ \rm{erg cm$^{-2}$ s$^{-1}$}}
  \sim800$ and $n_{\rm H}\sim2000$. For NGC4214-II, we find that 
   $G_{0}\sim1000$ and $n_{\rm H}\sim3000$.
The solution for the third
region NGC4214-III infers a lower incident
radiation field ($G_0\lesssim100$)}. 
The quoted values above take into
account the contribution to the [C~{\sc ii}] intensity of the
ionized medium between 0 and 40\%.
In all cases, these PDR parameters are
indicative of a relatively moderate gas density and radiation field. 
 
\subsection{[C~{\sc ii}]-to-CO ratio}
\label{sec:c-sc-ii}
We measure [C~{\sc ii}]/CO intensity ratios of 70\,000 and
20\,000 for NGC4214-I and II, respectively
(Table~\ref{table:table1}). These values are at least an order of
magnitude higher than those inferred for spiral or starburst galaxies
\citep{stacey-1991,negishi-2001} but in closer agreement with those of dwarf galaxies
\citep{poglitsch-1995,madden-1997,hunter-2001}. NGC4214-I has a
[C~{\sc ii}]/CO similar to 30 Dor \citep{ poglitsch-1995,rubin-2009}.
This high ratio could be explained by the paucity of dust; less dust shielding allows
the photodissociating FUV photons to penetrate deeper into the cloud,
reducing the size of the CO cores while leaving a large [C~{\sc
  ii}]-emitting envelope. This envelope may well contain self-shielded
molecular hydrogen. If this were true, the [C~{\sc ii}] would be a
more accurate tracer of the molecular reservoirs in low metallicity
environments than CO.
 
\section{Conclusion}
\label{sect:discussion}
These observations of NGC4214 with {\it Herschel}/{PACS} have enabled us to
spatially resolve the FIR fine-structure lines in the low metallicity
dwarf galaxy NGC4214.
A very striking result from the mapping is the brightness of the
[C~{\sc ii}]~158$\mu$m , [O~{\sc i}]~63$\mu$m, and the [O~{\sc
  iii}]~88$\mu$m lines. The [C~{\sc ii}] alone corresponds to 0.5\% of
the total infrared luminosity. The combined luminosity of the FIR lines
reaches 2\% of the L$_{\rm TIR}$. The [O~{\sc iii}] 88$\mu$m line is more
confined to star-forming regions and the [O~{\sc iii}]/[C~{\sc ii}]
reaches almost a factor of 2 close to the SSC in NGC4214-I.

The {\revised [C~{\sc ii}]/CO
ratios are also exceptionally high} (20\,000 $-$ 70\,000). 
These extreme ratios are indicative of an ISM that is strongly
affected by photodissociation and reflect the combined effects of 
low metallicity ISM and intense star formation. 
It should be borne in mind that, when comparing with models, 
the models predict line ratios for a
single, homogeneous PDR. In reality, we observe a multiphase
medium that {\revised probably harbors many extremely} different PDR conditions.
Therefore, this preliminary comparison indicates the need for more
complete (complex) models to be able to derive accurate physical
quantities. 

\begin{acknowledgements}
  We would like to thank {\revised the referee for suggestions that improved this letter}. 
  Many thanks to Fabian Walter for sharing his valuable CO
  data with us, and Leonardo \'{U}beda for his HST images for us to
  compare with. We are also extremely grateful for the help of the
  {PACS} ICC and {\it Herschel} Science Centre. {PACS} has been developed by a
  consortium of institutes led by MPE (Germany) and including UVIE
  (Austria); KU Leuven, CSL, IMEC (Belgium); CEA, LAM (France); MPIA
  (Germany); INAF-IFSI/OAA/OAP/OAT, LENS, SISSA (Italy); IAC (Spain).
  This development has been supported by the funding agencies BMVIT
  (Austria), ESA-PRODEX (Belgium), CEA/CNES (France), DLR (Germany),
  ASI/INAF (Italy), and CICYT/MCYT (Spain).
\end{acknowledgements}

\bibliographystyle{aa}
\bibliography{14699ref}

\end{document}